\documentclass{aa}
\usepackage{txfonts}
\usepackage{natbib}
\usepackage{graphicx}
\bibpunct{(}{)}{;}{a}{}{,}
\newcommand{\lin}{Sr~{\sc i}~$\lambda$4607~\AA}

\begin{document}
\title{On the \lin\
 Hanle depolarization signals in the quiet Sun}
   \author{J.~S\'anchez~Almeida}
    \institute{Instituto de Astrof\'\i sica de Canarias, 
              E-38205 La Laguna, Tenerife, Spain}
   \offprints{J.~S\'anchez Almeida, \email{jos@iac.es}}
   \date{Received~~~~~/ Accepted~~~~~}
\abstract{
The Hanle depolarization signals of 
\lin\ have been used to estimate the
unsigned magnetic flux and  magnetic energy
existing in the quiet Sun photosphere.
However, the \lin\ Hanle signals are  not sensitive 
to the unsigned flux and energy.
They only bear
information on the fraction of
photosphere occupied by magnetic field
strengths smaller than the Hanle saturation,
which do not contribute to the unsigned
flux and energy.
We deduce an approximate expression for
the relationship between magnetic fill factor and Hanle signal.
When applied to existing
Hanle depolarization measurements, it indicates that
only 40\% of the quiet Sun is filled by magnetic
fields with a strength smaller than 60~G. The
remaining 60\% of the surface 
has field strengths above this
limit. Such constraint will be needed to 
determine  the distribution of magnetic field strengths
existing in the quiet Sun.
\keywords{
	polarization --
	Sun: magnetic fields --
        Sun: photosphere 
	}
}
\maketitle
%

\section{Introduction}
Most of the solar surface does not show significant
polarization signals in the traditional magnetic field
measurements.
However, the magnetism of this so-called quiet Sun
shows up as soon as  the polarimetric sensitivity and 
the angular resolution exceed a threshold 
\citep[e.g.,][and references therein]{san04}.
Magnetic fields
in  the quiet Sun were discovered back in the
seventies by \citet{liv75} and \citet{smi75}, but  they
experience a revival due to the 
significant amounts of magnetic flux and energy
that the quiet Sun may store
\citep{ste82,zir87,yi93,san98c}.

The quiet Sun magnetic
fields\footnote{Often referred to as Inter-Network
fields, Intra-Network fields or simply IN fields.}
have inspired theoretical studies on their
origin \citep{pet93,cat99a} and  interplay with granular 
convection \citep{cat01,ste02,vog03},
as well as on the 
coupling with coronal magnetic structures
\citep{sch03b,goo04}. 
Unfortunately, we still
lack of a firm theoretical framework to describe
their nature. From an observational point of view,
our understanding has also improved during
the last years. We can assess the
complex magnetic topology
of these magnetic fields 
\citep{san96,sig99,san00,lit02,san03c}, the existence large amounts
of unsigned magnetic  flux and energy
\citep{san00,dom03a},
the variety of intrinsic field strengths 
\citep{fau95,bia99,lin99,san00,soc02}, and some
other properties \citep[see][]{san04}.
All these advances notwithstanding, the observational
characterization of the quiet Sun fields is unsatisfactory.
The complex topology of the fields makes all 
present measurements prone to severe bias, a
fact that has to be acknowledged and
fixed up before
developing a reliable  observational
picture of quiet Sun magnetism.

Due to the complexity  of the fields,
they have to be characterized in terms of
probability density functions  (PDFs).
The magnetic field strength PDF is particularly
useful. It
gives the fraction of solar photosphere occupied by
magnetic fields with a given field strength. The two 
first moments of this PDF provide the unsigned
magnetic flux density and the magnetic energy density, respectively 
\citep[e.g][]{san04}. From an observational point of view,
the PDF has to be assembled from various different measurements.
The polarization signals created by Zeeman effect
are most sensitive to magnetic field strengths larger
than a few hundred G. The Hanle effect induced signals
respond to field strengths smaller than this
limit. 

A significant part of what we know about the
weakest fields
comes from the interpretation of the Hanle 
depolarization signals of \lin\ 
\citep{ste82,fau93,fau95,ste97b,fau01}.
In particular,
it has been recently shown that the observed
Hanle depolarization signals of \lin\ require 
an average field of 130~G and a magnetic energy density
of $1.3\cdot 10^{3}$~erg~cm$^{-3}$ \citep{tru04}.  
However, this mean field exceeds
the limit able to induce significant
Hanle depolarization signal in this line 
($\sim$100~G; see, \S~\ref{basic} or \citealt{fau01}).
As we will discuss (\S~\ref{exponential}), this seeming 
inconsistency results from an assumption which turns out
to be decisive to assign a mean
field and  a magnetic energy to the observed
Hanle depolarization signals,
namely, the shape of the PDF.
Such PDF-dependence
of the inference 
casts doubts on
the results and, more importantly,
it  urges us to understand what is the true information
provided by \lin . Using a simplified (yet realistic) treatment of the
radiative transfer, we explore the diagnostic content under the 
conditions to be expected
 in the quiet solar photosphere, with a mean field
strength larger than the saturation field of the Hanle effect.
As a result, we find that the Hanle signals are sensible only
to the fill factor of magnetic fields below the Hanle saturation. 
They provide almost no constraint on the mean field strength and
magnetic energy that may exist in the quiet Sun.
This fact has to be taken into account  when using 
information from \lin\ to determine the PDF of the quiet
Sun magnetic field strength.

The work is organized as follows:
\S~\ref{basic} lists the approximations
used to carry out the Hanle depolarization
syntheses. The degree of realism
 of this treatment is discussed in
\S~\ref{realism}.
The Hanle signals to be expected
from  an exponential PDF with a mean field of
130~G are analyzed in
detail in \S~\ref{exponential}.
The actual diagnostic content of the
\lin\ Hanle signals is worked out in
\S~\ref{diagnose}.
An example of PDF with unbound magnetic flux and energy
compatible with the observed Hanle signals is
shown in \S~\ref{example}. 
Finally, the way 
in which these results constrain the 
distribution of magnetic fields existing 
in the quiet Sun are discussed in 
\S~\ref{conclusions}.

%
\section{Hanle depolarization of \lin}
\subsection{Basic properties and notation}\label{basic} 
\begin{figure}
\resizebox{\hsize}{!}{\includegraphics{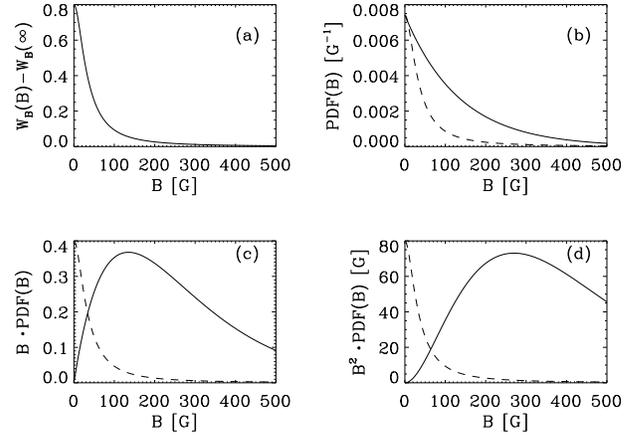}}
\caption{(a) \lin\ Hanle depolarization signals
versus magnetic field strength
for $B_H$ typical of the upper photosphere
of the quiet Sun. The depolarization signals
have no significant contribution when $B$ is, say, 
larger than 100~G.
(b) Exponential PDF with mean 130~G (the solid line).
(c) Unsigned
flux density per unit of magnetic field strength for the PDF shown
in (b). (d) Magnetic energy density per unit of 
magnetic field strength. The dashed lines in (b), (c) and
(d) represent $[W_B(B)-W_B(\infty)]$ scaled to
fit in the plots.
}
\label{hanle5a}
\end{figure}

\begin{figure}
\resizebox{\hsize}{!}{\includegraphics{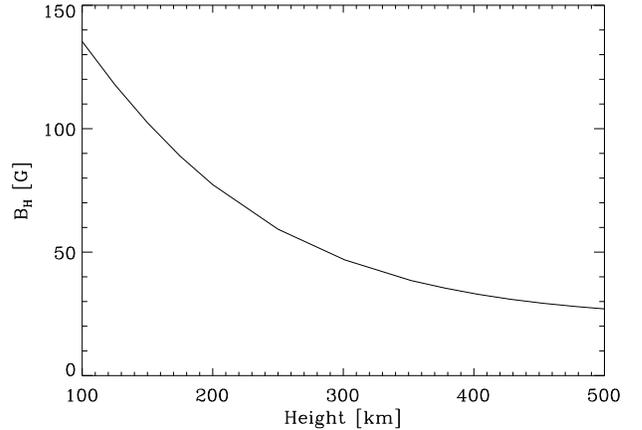}}
\caption{$B_H$ as a function of height in the atmosphere
for the quiet Sun model atmosphere by \citet{mal86}. Note that
$B_H$ is smaller than 50~G above 250~km,
where \lin\ is formed.
}
\label{fig1}
\end{figure}

The Hanle depolarization  of \lin\ 
can be expressed  as 
\begin{equation}
Q/Q_0\simeq W_B(B)=1-{2\over 5}\Big({{\gamma_H^2}\over{1+\gamma_H^2}}+
	{{4\gamma_H^2}\over{1+4 \gamma_H^2}}\Big),
	\label{hanle1}
\end{equation}
where 
$Q/Q_0$ is the ratio between the observed linear polarization
$Q$ and the polarization if there were no magnetic field $Q_0$. 
The symbol $\gamma_H$ parameterizes the magnetic field of the
microturbulent distribution of magnetic fields with random 
orientation and constant field strength $B$,
\begin{equation}
\gamma_H = B/B_H.
	\label{hanle2}
\end{equation}
The so-called Hanle parameter  $B_H$ scales linearly 
with the radiative transition rate plus the 
depolarizing collision rate, and it can be computed
from the temperature and density according to the
prescription in \citet[][\S~3]{fau01}.
The relationship (\ref{hanle1}) is an approximation
that holds when the Hanle depolarization 
results from a single scattering \citep{ste82,lan85b,fau01}.
It is a good approximation
for this particular line in the 
quiet Sun (see \S~\ref{realism} below).
The depolarization $W_B(B)$ does not vary significantly 
with $B$ when $B >> B_H$. This property is often referred to as
the saturation of the Hanle signal, implying that 
the Hanle depolarization is not sensitive to
fields much larger than $B_H$. 
Figure~\ref{hanle5a}a
shows $W_B(B)$ when $B_H=47~G$. Note that 
the depolarization barely changes for $B > 100$~G 
or $2 B_H$.
The value for $B_H$ used above is typical of \lin\
in the photospheric heights where the line is 
formed, say, from 200~km to 400~km above the base
of the photosphere \citep[][ which also show how 
the range of heights contributing to Hanle 
signals does not depend very much on the 
heliocentric angle]{fau95}.
Figure  \ref{fig1} shows $B_H$ as a function of height
in a  quiet Sun model atmosphere 
\citep{mal86}. 
$B_H$ is of the order of
$77~$G at 200~km and 33~G at 400~km, with a value of 47~G
in the middle of the range.
In the case that the magnetic field strength is not unique but
has a distribution of values according to a PDF $P(B)$,
then the  depolarization signal $<Q/Q_0>$ 
equals the average depolarization factor (see \citealt{lan85b}, \S 3),
\begin{equation}
<Q/Q_0>\simeq\int_0^\infty P(B)W_B(B)dB.
	\label{important}
\end{equation}
Since $<Q/Q_0>\not= 0$ when $B\rightarrow\infty$,
the actual range of polarization signals
sensitive to magnetic field strength variations
is only a part  of $<Q/Q_0>$, namely, 
\begin{equation}
<Q/Q_0>-<Q_\infty/Q_0>=\int_0^\infty P(B)[W_B(B)-W_B(\infty)]dB,
	\label{definition}
\end{equation}
where $Q_\infty$  is the linear polarization signal 
to be expected if only fields larger than
the  Hanle saturation are present in the atmosphere. For these
PDFs, $P(B)\not= 0$ only
when  $W_B(B)\simeq W_B(\infty)$, so that
 \begin{equation}
<Q_\infty/Q_0>\simeq\int_0^\infty P(B)W_B(\infty)dB=W_B(\infty)=1/5.
\label{infty_limit}
\end{equation}

Two particular cases are of special interest. They
have been used to assign  unsigned magnetic fluxes and
magnetic energies to observed \lin\ Hanle signals. In  the case that 
$P(B)$ is a very narrow function of $B$, namely,
\begin{equation}
P(B)\not= 0 {\rm ~only~when}~ |B-\widehat{B}| < \Delta B << B_H,
\end{equation}
then
\begin{displaymath}
<(Q-Q_\infty)/Q_0>\simeq\int_0^\infty P(B)
	[W_B(\widehat{B})-W_B(\infty)]dB
\end{displaymath}
\begin{equation}
	\simeq W_B(\widehat B)-W_B(\infty).
	\label{meanfield}
\end{equation}
The second case assumes $P(B)$ to be an exponential
function of mean $B_0$, 
\begin{equation}
P(B)=B_0^{-1}\exp(-B/B_0).
\end{equation}
Then 
equation~(\ref{definition}) admits 
a compact expression,
\begin{equation}
<(Q-Q_\infty)/Q_0>=
{2\over 5}\Big[{{B_H}\over{B_0}}f({{B_H}\over{B_0}})
	+{{B_H}\over{2B_0}}f({{B_H}\over{2B_0}})\Big],
	\label{mastereq}
\end{equation}
with 
\begin{equation}
f(x)=\int_0^\infty{{\exp(-xy)}\over{1+y^2}}dy.
\end{equation}

 \subsection{Accuracy of the single
 	 scattering approximation}\label{realism}
The equations described in the previous
section  hold for a single scattering event. However,
they are also a good
representation of the \lin\ Hanle signal when a more complete
treatment of the radiative transfer is considered.
\citet{fau01} compare full 1D NLTE 
syntheses with equation~(\ref{meanfield}),
and they conclude that both approaches
lead to the same $\widehat{B}$ within the error
bars of the observations. 
\citet{tru04} work out the \lin\ Hanle 
depolarization in
a realistic 3D hydrodynamic simulation, considering
a 15 level Sr~{\sc i} atom and solving the NLTE 
problem in three dimensions. Equations~(\ref{meanfield})
and (\ref{mastereq}) still provide magnetic fields
in close agreement with the full calculation.
The depolarization signal found by \citet{tru04} 
is 
\begin{equation}
<Q/Q_0>\simeq 0.41\pm0.04,
	\label{obs_depol}
\end{equation}
for $0.2 \leq \mu\leq 0.6$ ($\mu$ stands for the cosine
of the heliocentric angle).
In order to reproduce these observations, \citet{tru04}
need an exponential PDF with 
$B_0=130$~G or, alternatively,
a delta-function PDF with $\widehat B=60$~G. 
These values of $B_0$ and
$\widehat{B}$ also reproduce the observable~(\ref{obs_depol})
under the single scattering approximation.
Using $B_H=47~$G (which is the mean value 
in the atmospheric
layers  where the 
\lin\ Hanle depolarization is
formed), 
equation (\ref{mastereq}) with $B_0=130$~G leads to 
\begin{equation}
<Q/Q_0>\simeq 0.42.
	\label{observed_ratio}
\end{equation}
In addition, one obtains exactly the same depolarization
setting $\widehat B=57$~G in equation (\ref{meanfield}).

A comment is in order.
The large differences between the mean field 
strengths deduced by \citet[][$\widehat{B}\simeq$~25~G]{fau01} and 
\citet[][$\widehat{B}\simeq$~60~G]{tru04} 
are not due the different ways in which they synthesize the  
Hanle depolarization $<Q/Q_0>$ in terms 
of a turbulent magnetic field. This step of the 
estimate seems to be independent of the details
of the modeling, and  it
is properly described by equation~(\ref{important}).
The differences are
due to the estimate of the scattering signals
to be expected for no magnetic field (i.e., 
the quantity\footnote{The estimate by \citet{tru04}
is to be favored for quiet Sun diagnosis
since it is based on realistic MHD simulations, 3D scattering
polarization calculations,
and a complex Sr atom.}$<Q_0>$).
Although they start off from similar 
observed $<Q>$, \citet{fau01} infer $<Q/Q_0>\simeq 0.6$
instead of the depolarization in equation~(\ref{obs_depol}).
If
\citet{tru04} would have found 
$<Q/Q_0>=0.6$, their treatment would have
assigned 
$\widehat{B}\simeq$~30~G
to this signal. Such conclusion can be readily inferred from their
Fig.~1 by artificially increasing the observed $<Q>$
to yield $<Q/Q_0>\simeq 0.6$.
(See also \citealt{shc03}.)

\section{Largest field strength contributing to the 
Hanle signals}\label{exponential}
This section analyzes in detail the case of an
exponential PDF with $B_0=130$. It 
illustrates the difficulties to
infer unsigned magnetic fluxes and energies from the
quiet Sun Hanle depolarization signals of \lin\ .
 
\begin{figure}
\resizebox{\hsize}{!}{\includegraphics{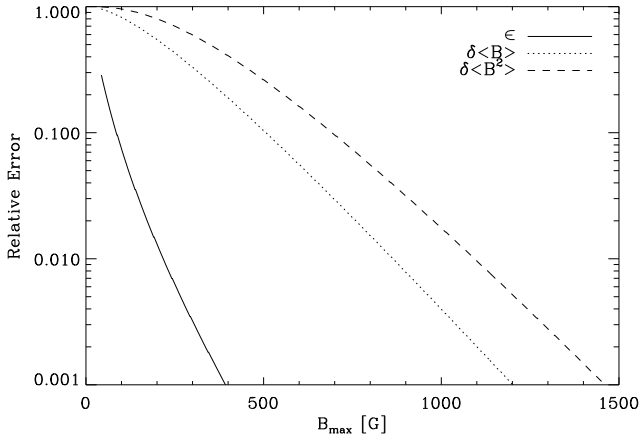}}
\caption{Relative errors of the Hanle
depolarization signals produced by neglecting 
all magnetic fields larger $B_{\rm max}$ 
(the solid line). It is smaller than the 10\% observational
error  when
$B_{\rm max}> 87$~G. Relative contribution to the
unsigned magnetic flux density (the dotted line) and
the magnetic energy density (the dashed line) produced
by magnetic fields larger than $B_{\rm max}$}.
\label{hanle5b}
\end{figure}

Let us define $B_{\rm max}$ as the largest field strength that
produces a significant contribution to the observed signal.
The contribution would be regarded as significant only if it
is above the observational error
$\epsilon$.
Then $B_{\rm max}$ is defined as
\begin{equation}
\epsilon=
{{\int_{B_{\rm max}}^\infty P(B)[W_B(B)-W_B(\infty)]dB}
\over
{\int_{0}^\infty P(B)[W_B(B)-W_B(\infty)]dB}},
\end{equation}
which  follows from equation (\ref{definition})
and the condition that all  magnetic fields with
 $B > B_{\rm max}$ change the
depolarization by less than a factor $\epsilon$. 
Figure \ref{hanle5b}, the solid line, shows $\epsilon=\epsilon(B_{\rm max})$
 when $B_H=47~G$ and
$B_0=130~G$, i.e., the parameters that characterize the observed
depolarization. 
The true observational error is larger than 10\% (see \S~\ref{realism}). 
Note that 
$\epsilon < 0.1$  when $B_{\rm max} \geq 100$~G,
meaning that those magnetic fields of the exponential
PDF with $B> 100~$G do not contribute to the
observed signal.

The unsigned magnetic flux density and magnetic energy density
can be computed from the first two moments
of the PDF \citep[e.g.][]{san04}.
However, the magnetic fields that determine
the first two moments of an exponential
PDF with $B_0=130~$G exceed  $B_{\rm max}\simeq$100~G.
Consequently, the unsigned flux and
the energy  assigned assuming an exponential PDF
are not constrained by the observations. 
The first two moments are given by
\begin{equation}
<B> =\int_0^{\infty} B\cdot P(B) dB
	\label{meanb},
\end{equation}
and
\begin{equation}
<B^2> =\int_0^{\infty} B^2\cdot P(B) dB.
\label{fluxb}
\end{equation} 
The integrands of equations (\ref{meanb}) and 
(\ref{fluxb}) describe the contribution of each
field strength to the two moments, and they are
shown in  Figs.~\ref{hanle5a}c and  \ref{hanle5a}d.
The figures include a scaled version of $[W_B(B)-W_B(\infty)]$
(the dashed lines), which indicates
the range of field strengths constrained
by the observations. 
Obviously,
most of the unsigned flux (related to the first moment) and 
energy density (second moment) are provided by field
strengths larger than ~100~G.
It is possible to quantify the bias
defining the fraction of unsigned flux density
made up by considering field strengths
which do not contribute to the Hanle signals,
explicitly, 
\begin{equation}
\delta{<B>} ={{\int_{B_{\rm max}}^\infty B\cdot P(B) dB}
	\over{\int_0^{\infty} B\cdot P(B) dB}}.
\end{equation}
Similarly, the fraction of energy density that 
comes from the tail of the PDF,  and so, it does not
contribute to the Hanle signals, is 
\begin{equation}
\delta{<B^2>} ={{\int_{B_{\rm max}}^\infty B^2\cdot P(B) dB}
	\over{\int_0^{\infty} B^2\cdot P(B) dB}}.
\end{equation}
Both $\delta<B>$ and $\delta<B^2>$ are shown in 
Fig~\ref{hanle5b}. It turns out
that  90\% of the Hanle signals ($\epsilon=0.1$) are
produced by $B<86$~G.  The tail of magnetic fields 
which do not contribute to the Hanle signals actually
produce 86\% of $<B>$ and 97\% of $<B^2>$. 
In other words,  the result by \citet{tru04}
that $<B>=B_0=130~$G
and $<B^2>/(8\pi)=\ (2B^2_0)/(8\pi)=\ 1.3\cdot~10^{3}~$erg~cm$^{-3}$ 
is based on the
assumption of the shape of PDF, but it is  not 
constrained by the observations. By changing the 
tail of large field strengths of the PDF one can modify
in an arbitrary manner the unsigned flux and
energy, yet producing the observed 
Hanle depolarization. A bold example is given 
in~\S~\ref{example}

\section{What does the Hanle depolarization
of \lin\ diagnose?}\label{diagnose}

Let us consider PDFs with the properties to be expected
for the solar quiet Sun magnetic fields.
The Hanle signals
are produced by magnetic fields 
within the bandpass where $[W_B(B)-W_B(\infty)]\not=0$,
which is of the order of  $2B_H$.
Then the Hanle signals can be
estimated as
\begin{equation}
<(Q-Q_\infty)/Q_0>\simeq \int_0^{2B_H}
P(B)[W_B(B)-W_B(\infty)]dB.\label{trickti}  
\end{equation}
In addition, the quiet Sun PDF should 
spread out over
a range of field strength
larger than $B_H$ \citep[see, e.g.,][]{san04}, so that
within the bandpass of interest,
it can be approximated by a linear expansion, 
\begin{equation}
P(B)\simeq a+b (B-B').
\end{equation}
Obviously,
\begin{equation}
a={1\over {2B'}}\int_0^{2B'}P(B)dB.
	\label{eq1}
\end{equation}
If one chooses $B'$ to be
\begin{equation}
B'={{\int_0^{2B_H} B 
	[W_B(B)-W_B(\infty)]dB}\over{\int_0^{2B_H}  
	[W_B(B)-W_B(\infty)]dB}},
	\label{bmaxdef}
\end{equation}
then the Hanle depolarization signals given by
equation (\ref{trickti}) are  independent
of the slope $b$, 
\begin{equation}
<(Q-Q_\infty)/Q_0>=C~
	\int_0^{2B'}P(B)dB,
\end{equation}
\begin{displaymath}
	C={1\over {2B'}}\int_0^{2B_H}
	[W_B(B)-W_B(\infty)]dB.
\end{displaymath}
The integrals defining $B'$ and $C$ can be solved
analytically to render 
\begin{equation}
B'\simeq 0.65~B_H,~~
C\simeq 0.54,
\end{equation}
and, consequently,
\begin{equation}
<(Q-Q_\infty)/Q_0>
\simeq 0.54 \int_0^{1.3B_H} P(B)dB.
	\label{trick}  
\end{equation}
The Hanle depolarization signals 
scale with the  fill factor of fields
with strength smaller than $1.3B_H$. As
expected, the Hanle depolarization of \lin\ is independent
of the actual unsigned magnetic flux or magnetic
energy of the distribution. 

\begin{figure}
\resizebox{\hsize}{!}{\includegraphics{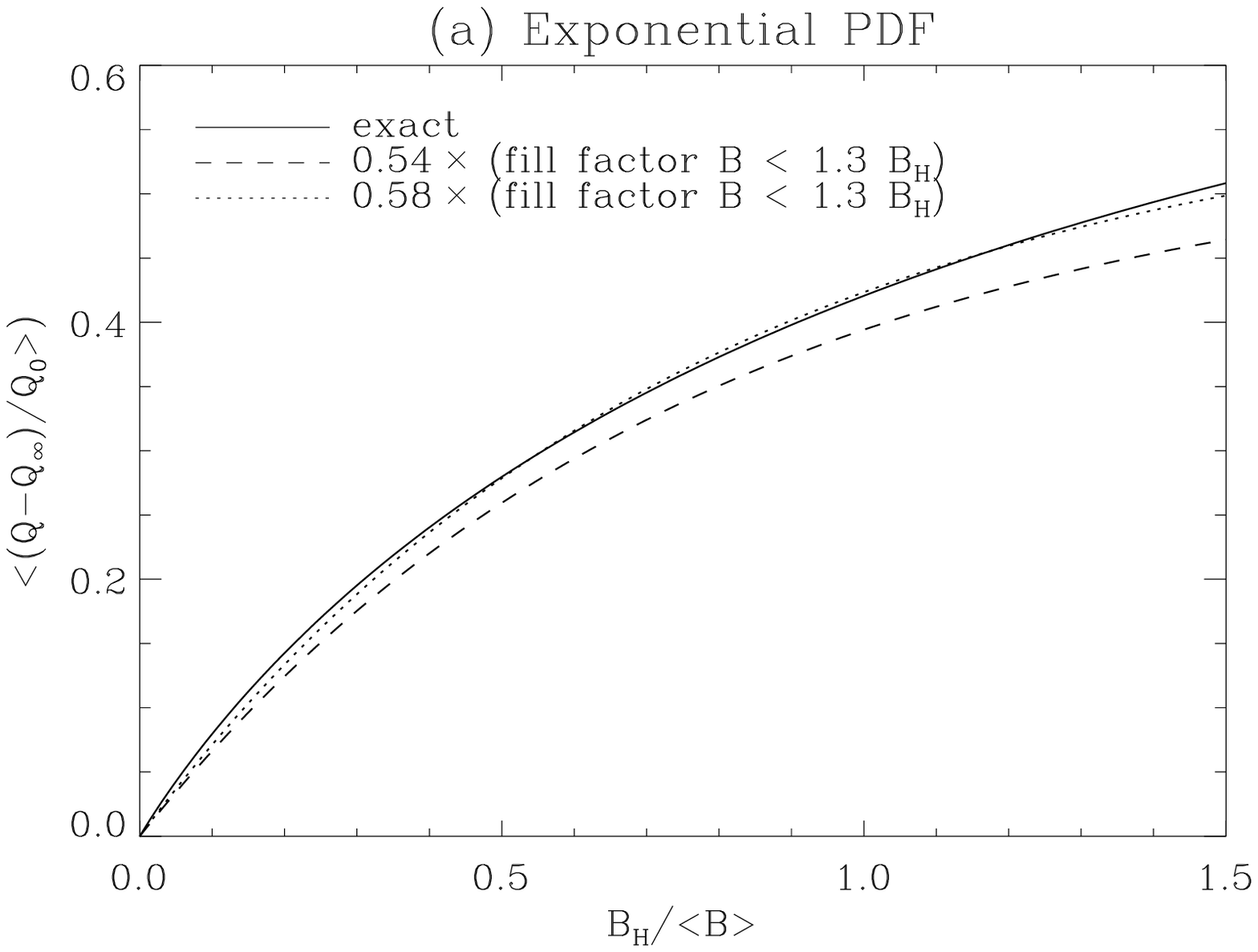}}
\resizebox{\hsize}{!}{\includegraphics{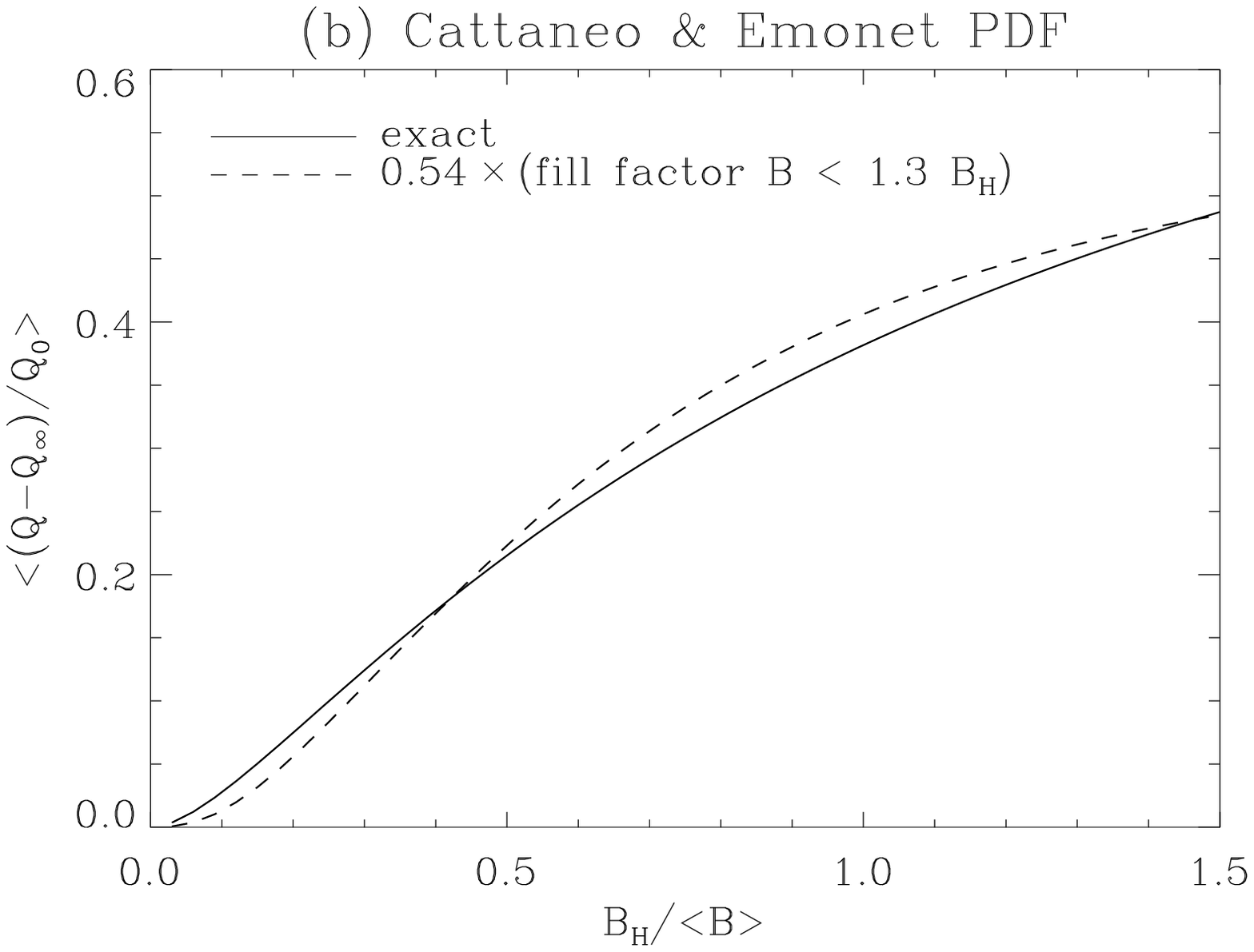}}
\label{hanle7}
\caption{
(a)
Hanle depolarization as 
a function of $B_H/<B>$ when the 
PDF is an exponential function (the solid line).
It is approximately given by half the fill factor
of magnetic fields smaller than $1.3 B_H$ (the dashed
line). 
(b) Hanle depolarization as 
a function of $B_H/<B>$ for the PDF of the 
turbulent dynamo simulations by 
\citet{emo01}, as described by  \citet{san03} (the solid
line). The approximate relationship worked out in the
paper also provides a fair representation
of the Hanle signals.
}
\end{figure}

The approximate  equation~(\ref{trick}) has been
tested for the case
where $P(B)$ is exponential. 
Figure~\ref{hanle7}a shows the variation of the Hanle
depolarization signals with $B_H/<B>$, where $<B>$
stands for 
the mean magnetic field of the distribution.
The solid line has been computed from the exact solution~(\ref{mastereq}),
whereas the dashed line follows from the approximate
equation~(\ref{trick}). The agreement
is good, and it improves to excellent if $C=0.58$ (the dotted line). 
Figure \ref{hanle7}b also shows the exact and the approximate
Hanle signals deduced for the numerical turbulent
dynamo  PDF used
in \citet{san03}. The agreement of the approximation is also
satisfactory.
Despite the simplistic approximation leading to the
relationship~(\ref{trick}), it reproduces within 
10\%
the exact relationship~(\ref{definition}). The agreement is
expected independently
of the (unknown)  details of the
quiet Sun PDF, since equation~(\ref{trick}) it is based on
very general assumptions.

Equation~(\ref{trick}) satisfies the observation~(\ref{obs_depol})
when magnetic field strengths smaller than
60~G occupy 40\% of the quiet Sun. Obviously,  the 
exponential PDF chosen by \citet{tru04} fulfills
this criterion. The turbulent dynamo PDF used
by \citet[][\S~5]{san03} does it too.

\section{Example of unbound magnetic flux and
energy compatible with the observed Hanle 
signals}\label{example}
	From the arguments spelled out
above, it is clear that the unsigned magnetic flux and energy
of the distribution are not related
to the Hanle signals. This section  shows and example
reinforcing the point. We show
a  PDF producing the observed polarization with an
arbitrarily large magnetic flux and energy. Consider
the combination of exponentials,
\begin{equation}
P(B)=\beta B_1^{-1}\exp(-B/B_1) + (1-\beta) B_2^{-1}\exp(-B/B_2),
\label{pdf2}
\end{equation}
with $B_1<< B_H << B_2$. The second
exponential do not contribute to the Hanle depolarization
signals, since it is given by equation~(\ref{mastereq}) 
with $B_H/B_0\rightarrow 0$. On the other hand,
$f(x)\rightarrow x^{-1}$ when $x\rightarrow \infty$, which
together with equations~(\ref{mastereq}) and
(\ref{infty_limit})  lead to 
\begin{equation}
<Q/Q_0>\simeq (1+4\beta)/ 5.
	\label{approx_depol}
\end{equation}
The value $\beta\simeq 0.27$ satisfies the observed
depolarization~(\ref{obs_depol}).
On the other hand, the
two moments~(\ref{meanb}) and (\ref{fluxb}) only depend on the second 
component,
\begin{equation}
<B>=\beta B_1 + (1-\beta) B_2\simeq (1-\beta) B_2,
\end{equation}  
\begin{equation}
<B^2>=2\beta B_1^2 + 2(1-\beta) B^2_2\simeq 2(1-\beta) B^2_2.
\end{equation}
since $B_1 << B_2$. Then the unsigned flux and the
energy corresponding to equation~(\ref{pdf2}) 
can be arbitrarily large by increasing $B_2$ with no
modification of the Hanle signals (equation [\ref{approx_depol}]).

\section{Conclusions}\label{conclusions}
The Hanle depolarization signals of 
\lin\ are insensitive to the magnetic
flux and magnetic energy existing in the
quiet Sun photosphere. They only bear
information  on fraction of surface
occupied by magnetic field
strengths smaller than some $2 B_H$,
equivalent to  100~G
when the so-called Hanle parameter $B_H$ is $50$~G.
In order to gain
physical insight into the diagnostic  
provided by the
\lin\ Hanle depolarization signals,
we estimate the signals to be expected
when the magnetic atmosphere 
contains disorganized
magnetic fields with field strengths
spanning from zero to a value exceeding $B_H$. 
The expected signal
is given by equation~(\ref{trick}).
It is a fixed fraction (54\%)
of the fill factor of magnetic fields whose
strengths are smaller than $1.3~B_H$.

Our result implies that the findings by 
\citet{tru04} have to be updated, in particular,
the Hanle depolarization signals that they work out
(equation~[\ref{observed_ratio}])
do not imply an unsigned flux  and
a magnetic energy density.
They imply that  
40\% of the quiet Sun is covered by 
fields whose strength is smaller than 60~G. In 
other words, some 60\% of the quiet
Sun has field strengths
larger than 60~G. Obviously, this result
is consistent with a distribution 
of magnetic fields  having
a magnetic flux and energy even larger than 
those inferred by \citet{tru04}, but 
the unsigned flux and 
energy are not constrained by the Hanle signals.
The magnetic flux and 
energy are constrained by the 
polarization signals induced via  
Zeeman effect,  which
contain information on the 
hG and kG field strengths of the quiet Sun PDF \citep[e.g.,][]{san00}.
For example, the magnetic energy density of the
kG fields observed by \citet{dom03a}, which occupy
only 2\% of the surface, actually 
contain as much energy as an
exponential PDF with $B_0=$130~G.

We  lack of a reliable quiet Sun PDF to represent
the whole range of observed field
strengths from zero to two
kG. This PDF would have to be assembled piecing together
information from various sources, as it is indicated
in \citet{san04}. The results of our
analysis are needed to complete
such observational work, since they clarify how
the \lin\ Hanle signals constrain the empirical PDF.
%
%

\acknowledgements{
%
%
The work has partly been funded by the Spanish Ministry of Science
and Technology, 
project AYA2004-05792, as well as by
the EC contract HPRN-CT-2002-00313.
}

%

\end{document}